\documentclass{IEEElmag}

\usepackage[colorlinks,urlcolor=blue,linkcolor=blue,citecolor=blue]{hyperref}

\usepackage[hyphenbreaks]{breakurl}
\usepackage{physics}
\usepackage{color,array}

\usepackage{graphicx}

\usepackage{CJKutf8}

\begin{document}

\sptitle{Magnetic tunnel junction equivalent circuit modeling}

\title{A Physics-Based Circuit Model for Magnetic Tunnel Junctions}

\author{Steven Louis\affilmark{1}*}
\author{Hannah Bradley\affilmark{2}} 
\author{Artem Litvinenko\affilmark{3}}
\author{Vasyl Tyberkevych\affilmark{2}}

\affil{Department of Electrical and Computer Engineering, Oakland University, Rochester, MI, 48309, USA}
\affil{Department of Physics, Oakland University, Rochester, MI, 48309, USA}
\affil{Department of Physics, University of Gothenburg, Fysikgr\"{a}nd 3, Gothenburg, 412 96, Sweden}

\corresp{Corresponding author: Steven Louis (slouis@oakland.edu).}


\begin{abstract}
This work presents an equivalent circuit model for magnetic tunnel junctions (MTJs) that accurately reproduces their magnetization dynamics and electrical behavior within the macrospin approximation.
The model is validated through direct numerical simulations of the Landau–Lifshitz–Gilbert–Slonczewski (LLGS) equation, encompassing ferromagnetic resonance, field- and spin-torque-induced switching, and spin-torque-induced oscillations.
Simulation results exhibit excellent agreement between the equivalent circuit model and the LLGS-based simulations, confirming the model accuracy and utility for efficient circuit-level analysis of MTJs.
The capability of handling time-dependent magnetic fields and voltage-driven excitations renders the proposed model applicable to diverse areas, including neuromorphic computing, microwave signal processing, and spintronic memory technologies.
By providing a computationally efficient yet physically rigorous circuit representation, this work facilitates seamless integration of MTJs into complex electronic systems, thereby accelerating the advancement of novel spintronic circuit architectures.
\end{abstract}


\maketitle

\section{INTRODUCTION}\label{introduction}

In the past two decades, magnetic tunnel junctions (MTJs) have revolutionized data storage by enabling high-density recording in disk-based drives {[}Chappert, 2007{]}.
MTJs have also been integrated with CMOS technology and commercialized in magnetoresistive random-access memory (MRAM) {[}Alzate 2015, Edelstein 2020{]} which leverages the non-volatility and fast switching properties of MTJs for energy-efficient, high-speed memory {[}Ikegawa 2020{]}.
Beyond digital applications, MTJs may also be used as microwave sources {[}Fuchs 2004, Villard 2009{]}, microwave detectors {[}Tulapurkar 2005{]}, energy harvesters {[}Fang 2019{]}, advanced microwave signal processing devices {[}Litvinenko 2022{]}, and even in neuromorphic computing {[}Romera 2018, Louis 2024, Rodrigues 2023{]}.

An MTJ consists of two ferromagnetic layers separated by a thin insulating barrier {[}Singh 2019, Yuasa 2007{]}.
The operational principle of an MTJ is based on the tunneling magnetoresistance effect, which is sensitive to the relative orientation of magnetic moments in the ferromagnetic layers {[}Yuasa 2007{]}, and on the spin-transfer-torque (STT) effect {[}Slonczewski 1989, Ralph 2008{]} that describes influence of electric current on magnetization dynamics.
Advanced nanoscale fabrication enabled integration of MTJs into complex electronic systems, expanding their use in memory and logic devices {[}Alzate 2015, Edelstein 2020{]}.
These distinctive properties position MTJs as important components in next-generation analog and neuromorphic computing systems {[}Incorvia 2024{]}.

\begin{figure*}[t]
\centerline{\includegraphics{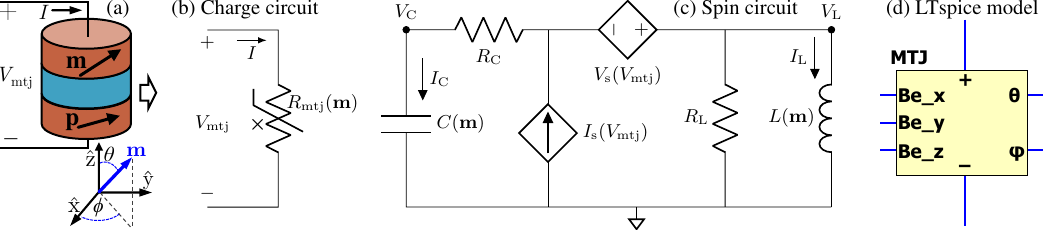}}
\caption{(a) Diagram of the MTJ, illustrating the free and fixed layers. 
(b) Equivalent circuit of the MTJ, as an intrinsic variable resistor.
(c) Equivalent circuit representation of the MTJ free layer magnetization.
(d) The complete MTJ circuit implemented in LTspice.
}
\label{set}
\end{figure*}

Circuit simulation is essential in the design and analysis of electronic systems {[}Najm 2010{]}.
An accurate circuit-equivalent model of an MTJ facilitates its integration into larger circuits and allows comprehensive performance evaluation, while preserving insight into magnetization dynamics. 
Prior works {[}Xu 2012, Panagopoulos 2012, Kim 2015{]} have used LTspice and other circuit simulators to numerically solve the LLGS equation, for example by mapping angular variables onto circuit elements. However, these approaches do not establish a physical correspondence between the circuit elements and the underlying energy storage, dissipation, or torque mechanisms present in the MTJ.
There are models {[}Bendjeddou 2023, Abbasi 2017{]} that can emulate the MTJ impedance at high frequency, but do not take into account MTJ behavior under the influence of static inputs.

This letter introduces an equivalent circuit model for MTJs that accurately reflects their underlying physical characteristics. 
Within the proposed model, conservative magnetic interactions (e.g., effect of a bias magnetic field or magnetic anisotropy) are modeled by equivalent nonlinear capacitance and inductance. 
Dissipative magnetic torques are represented by equivalent resistance (Gilbert damping) and current and voltage sources (STT). 
The model is realized as a standalone element in LTspice circuit simulation software and its validity is demonstrated by comparing LTspice simulation results with direct numerical simulations of the Landau-Lifshits-Gilbert-Slonczewski (LLGS) equation.


\section{Physics-Based Equivalent MTJ Circuit}

An MTJ comprises a ferromagnetic fixed layer, a non-magnetic spacer, and a ferromagnetic free layer, as shown in Fig.~\ref{set}(a).
The magnetization of the fixed layer is represented by $\mathbf{p}$, while the free layer magnetization is denoted by $\mathbf{m}$.
The term ``free layer'' indicates that the magnetization direction can be altered by external stimuli, such as an applied magnetic field $\mathbf{B}_\mathrm{ext}$ or electric current $I$.
In contrast, the fixed layer has a stable magnetization direction that is largely unaffected by external perturbations.
The dynamic behavior of $\mathbf{m}$ arises from the interplay between magnetic fields, damping effects, and spin-transfer torque, producing complex magnetization dynamics.
These dynamics determine the electrical behavior and switching characteristics of MTJ-based devices {[}Slavin 2009{]}.

When an electric current $I$ passes through the MTJ, it encounters a resistance denoted $R_\mathrm{mtj}$.
The resistance $R_\mathrm{mtj}$ reaches a minimum value, $R_\mathrm{P}$, when the magnetization vectors $\mathbf{p}$ and $\mathbf{m}$ are parallel.
Conversely, the resistance is maximized, taking the value $R_\mathrm{AP}$, when $\mathbf{p}$ and $\mathbf{m}$ are antiparallel.
If the vectors $\mathbf{p}$ and $\mathbf{m}$ are orthogonal, the resistance becomes $R_\perp$, defined by $R_\perp = 2 R_\mathrm{AP} R_\mathrm{P}/(R_\mathrm{AP} + R_\mathrm{P})$.
A general expression for the resistance of an MTJ is {[}Slonczewski 1989, Wang 2009{]},
\begin{equation} 
R_\mathrm{mtj}(\mathbf{m}) = \frac{R_\perp}{1 + \eta^2 (\mathbf{p} \cdot \mathbf{m}) }, 
\label{rmtj} 
\end{equation} 
where $\eta$ is the dimensionless spin polarization efficiency, given by $\eta = \sqrt{(R_\mathrm{AP} - R_\mathrm{P})/(R_\mathrm{AP} + R_\mathrm{P})}$. 
The MTJ shown in Fig.~\ref{set}(a) can be represented by the circuit in Fig.~\ref{set}(b), which incorporates an intrinsic magnetoresistor symbol as defined in {[}IEEE Standard{]} 2.1.7.
The resistance $R_\mathrm{mtj}(\mathbf{m})$ is intrinsically nonlinear, depending on the free layer magnetization $\mathbf{m}$.

The magnetization dynamics of the free layer can be effectively described using the macrospin approximation, which can be modeled by the Landau-Lifshitz-Gilbert-Slonczewski (LLGS) equation {[}Slavin 2009{]}, given by 
\begin{equation}
\dv{\mathbf{m}}{t} - |\gamma|\mathbf{B}_\mathrm{eff} \times \mathbf{m} = \alpha_G \mathbf{m} \times \dv{\mathbf{m}}{t} - \frac{\gamma}{M_s V_o}\mathbf{m} \times \mathbf{J}_s \times \mathbf{m}, 
\label{llgs}
\end{equation} 
where $\gamma$ is the gyromagnetic ratio, and $\alpha_G$ is the Gilbert damping parameter. 
The spin current $\mathbf{J}_s$ can be written as $\mathbf{J}_s = (\gamma\hbar\eta/2 e M_s V_o)(V_\mathrm{mtj}/R_\perp)\mathbf{p}$, where $\hbar$ is the reduced Planck constant, $\mu_0$ is the vacuum magnetic permeability, $e$ is the elementary charge, $V_o$ represents the volume of the free layer, and $V_\mathrm{mtj}$ is the voltage across the MTJ. 
The effective magnetic field $\mathbf{B}_\mathrm{eff}$ is given by $\mathbf{B}_\mathrm{eff} = -(1/M_s V_o)\partial E(\mathbf{m})/\partial\mathbf{m} $, where $E(\mathbf{m})$ is the macrospin energy which is given by 
\begin{equation}E(\mathbf{m}) = M_s V_o\Big[ -\mathbf{B}_e \cdot \mathbf{m} + \tfrac{1}{2} B_d (\mathbf{m}\cdot\vu{z})^2 - \tfrac{1}{2} B_a (\mathbf{m} \cdot \vu{x})^2\Big], \label{energy}\end{equation}
where $\mathbf{B}_e$ is the externally applied field, $B_{a}$ is the easy-axis anisotropy field in the MTJ plane, and $B_{d}$ is the effective demagnetizing field having contributions from dipolar field and perpendicular magnetic anisotropy (PMA). 
Please note that left side of (\ref{llgs}) represents conservative precession, while the right side includes Gilbert damping torque and Slonczewski spin-transfer torque (STT).

Equation~(\ref{llgs}) models the magnetization dynamics of the free layer and can be used to determine the macrospin vector $\mathbf{m}(\theta,\phi)$ as a function of time, where $\theta$ and $\phi$ are the polar and azimuthal angles of $\mathbf{m}$ (see Fig.~\ref{set}(a)).
This dynamic behavior can be equivalently modeled by a nonlinear LC circuit by defining an effective capacitor charge $Q$ and inductor flux $\Phi$
\begin{equation}
Q = \frac{e}{\hbar}\frac{M_s V_o}{\gamma} \cos\theta, \hspace{4mm} \Phi = \frac{\hbar}{e} \phi. 
\label{PQ}
\end{equation}

The voltage $V_C$ across a capacitor and the current $I_L$ through an inductor can be related to the energy $E$ of an LC circuit by
\begin{equation} V_C =\pdv{E}{Q}, \hspace{4mm} I_L = \pdv{E}{\Phi}.
\end{equation}
For an MTJ with an energy function (\ref{energy}), these definitions give
\begin{equation}V_C = \frac{\gamma \hbar}{e}\Big[B_{e,z}-\frac{(B_{e,x}\cos\phi + B_{e,y}\sin\phi)}{\tan\theta} -(B_d + B_a\cos^2 \phi)\cos\theta \Big],\label{cs1}\end{equation}
\begin{equation}I_L = \frac{\gamma \hbar}{e Z_s}\Big[ (B_{e,x}\sin\phi - B_{e,y}\cos\phi)\sin\theta + \frac{1}{2}B_a\sin(2\phi)\sin^2\theta \Big].\label{cs2}\end{equation}
Here, $Z_s = \hbar^2\gamma/(e M_s V_o)$ is a ``characteristic impedance'' of the macrospin. 
Equations (\ref{cs1}) and (\ref{cs2}) represent constitutive relations for equivalent macrospin capacitance and inductance; that is, they define how $V_C$ and $I_L$ relate directly to the magnetic state variables ($\theta$ and $\phi$), thereby providing the bridge between magnetization dynamics and circuit-level electrical behavior. 

From the constitutive relations (\ref{cs1}) and (\ref{cs2}), a capacitance $C$ can determined as $C = \left(\partial^2 E/\partial Q^2\right)^{-1}$, or
\begin{equation}C = \frac{1}{\gamma Z_s} \Big[\frac{1}{B_d + B_a \cos^2\phi +( B_{e,x} \cos\phi + B_{e,y} \sin\phi)/ \sin^3\theta}\Big]. \end{equation}
Likewise, an inductance $L= \left(\partial^2 E/\partial \Phi^2\right)^{-1}$ can be determined,
\begin{equation}L = \frac{Z_s}{\gamma} \Big[\frac{1}{(B_{e,x} \cos\phi + B_{e,y} \sin\phi ) \sin\theta + B_a \cos2\phi \sin^2\theta}\Big],\end{equation}
Please note that both the capacitance and inductance depend explicitly on the external magnetic field and the free layer magnetization angles $\theta$ and $\phi$.
For notational simplicity, we will express them as functions of the free layer magnetization: $C \rightarrow C(\mathbf{m})$ and $L \rightarrow L(\mathbf{m})$.

Fig.~\ref{set}(c) presents the equivalent circuit model that incorporates $C(\mathbf{m})$ and $L(\mathbf{m})$ along with additional elements:
\begin{equation}R_C = \frac{ \alpha_G Z_s }{\sin^2\theta}, \hspace{4mm} R_L = \frac{ Z_s }{ \alpha_G \sin^2\theta } , \end{equation}
\begin{equation}I_s(V_\textrm{mtj}) = \frac{\eta}{2}\Big[p_z\sin\theta - ( p_x\cos\phi + p_y\sin\phi )\Big]\sin\theta\frac{V_{\textrm{mtj}} }{R_\perp},\end{equation}
\begin{equation}V_s(V_\textrm{mtj}) = \frac{\eta}{2}\Big[\frac{p_x\sin\phi - p_y\cos\phi}{\sin\theta} \Big] \frac{V_{\textrm{mtj}} }{R_\perp} Z_s .\end{equation}

A complete derivation of these expressions and full development of the equivalent circuit (Fig.~\ref{set}(c)) is provided in {[}Louis 2025{]}. 

This equivalent circuit accurately captures the complete dynamics of $\mathbf{m}$.
In direct correspondence with (\ref{llgs}), the capacitor $C(\mathbf{m})$ and inductor $L(\mathbf{m})$ model conservative precession, resistors $R_C$ and $R_L$ represent Gilbert damping, and the sources $I_s(V_\textrm{mtj})$ and $V_s(V_\textrm{mtj})$ include MTJ voltage dependence, capturing STT effects.

This paragraph outlines the procedure for simulating the MTJ dynamics using the equivalent circuit model. 
The simulation begins by applying the inputs $V_\textrm{mtj}$ and $I$ to the charge circuit in Fig.~\ref{set}(b), while specifying the external magnetic field $\mathbf{B}_e$.
These inputs drive the spin circuit shown in Fig.~\ref{set}(c), which outputs the capacitor current $I_C$ and the inductor voltage $V_L$. 
According to standard circuit definitions, these are related to the state variables by 
\begin{equation} I_C = \dv{Q}{t}, \hspace{4mm} V_L = \dv{\Phi}{t}. \label{IVPQ} \end{equation} 
From $I_C$ and $V_L$, the circuit determines $Q$ and $\Phi$, which are then used in Eq.~(\ref{PQ}) to compute the free layer magnetization angles $\theta$ and $\phi$. 
This yields the magnetization vector $\mathbf{m}(\theta, \phi)$, allowing the MTJ resistance $R_\mathrm{mtj}(\mathbf{m})$ to be evaluated through Eq.~(\ref{rmtj}). 
Finally, this resistance is fed back into the charge circuit in Fig.~\ref{set}(b), completing one self-consistent iteration of the simulation.

The circuits shown in Fig.~\ref{set}(b) and Fig.~\ref{set}(c) were implemented in LTspice, a free circuit simulation software provided by {[}Analog Devices{]}. 
To simplify simulation workflows, these circuits were encapsulated into a single LTspice component, illustrated in Fig.~\ref{set}(d). 
This standalone component features input and output terminals labeled ``$+$'' and ``$-$'' for electrical signals, along with additional input ports for specifying the external magnetic field $\mathbf{B}_e$. 
The output ports $\theta$ and $\phi$ can be used to monitor internal state of the MTJ.
The complete LTspice model, including simulation examples and documentation, is available at {[}MTJ Spice Models{]}.

The equivalent circuit framework is designed to be modular and easily extendable. 
For instance, incorporating additional physical effects such as field-like torque involves modifying the macrospin energy expression in (\ref{energy}), which in turn alters the constitutive relations in (\ref{cs1}) and (\ref{cs2}). 
Similarly, configurations with more than one fixed magnetic layer, as explored in~{[}Louis2024{]}, can be modeled by extending Eq.~(\ref{rmtj}) to include multiple spin polarization terms and updating the spin circuit in Fig.~\ref{set}(c) with additional dependent sources. 
These examples highlight the flexibility of the model, making it well-suited for simulating a wide range of spintronic device architectures. 
These extensions follow naturally from the underlying energy-based formulation, and a comprehensive derivation of the equivalent circuit model is provided in {[}Louis 2025{]}.

Although this work focuses on a single macrospin, spatially non-uniform phenomena, such as spin-wave excitations and domain formation, could be approximated by networks of coupled macrospin circuit elements.  
This generalization offers a promising route to incorporate micromagnetic effects into circuit-level simulations.

\section{Validation of Equivalent Circuit Model}

To assess the accuracy and practical utility of the proposed model, we performed a series of simulations across representative MTJ operating regimes.
The simulated behaviors include ferromagnetic resonance (FMR), spin-torque nano-oscillator (STNO) operation, field-induced switching, and spin-torque switching.

\subsection{Ferromagnetic Resonance}

The FMR response was evaluated by comparing LTspice simulations of the equivalent circuit model with theoretical predictions. 
The simulations modeled free decay of magnetization oscillations, i.e., the magnetization was initialized slightly off equilibrium and then relaxed with oscillations at the FMR frequency.
Simulations used a free layer with magnetization $M_s = 796$~kA/m ($B_d = \mu_0M_s = 1$~T), volume of $V_o = 5.65 \times 10^{-24}$~m$^3$, corresponding to an elliptical geometry of the free layer with a semi-axes of 30~nm and 20~nm and a thickness of 3~nm, and the anisotropy field $B_a = 0.2$~T. 
To precisely measure the FMR frequency, the Gilbert damping parameter was set to a small value of $\alpha_G = 0.0001$.

\begin{figure}[t]
\centerline{\includegraphics{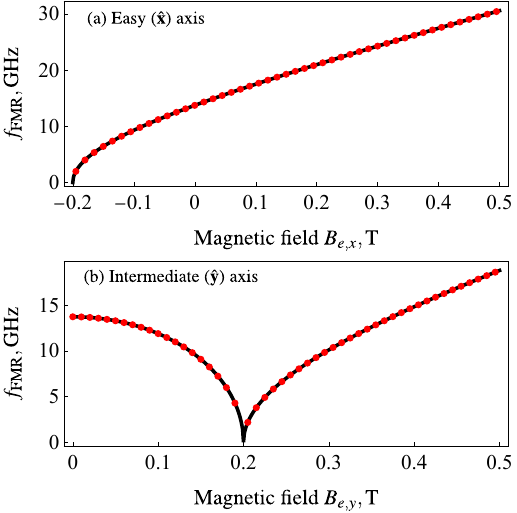}}
\caption{Comparison of simulated and theoretical FMR frequencies as functions of applied magnetic fields.  
The red dots represent LTspice simulations, while the black curves correspond to analytical predictions derived from the LLGS equation. External magnetic field is applied along (a) the easy axis $\vu{x}$ and (b) the intermediate axis $\vu{y}$.
}
\label{fmrG}
\end{figure}

Figure~\ref{fmrG}(a) shows that LTspice simulations (red dots) and theoretical FMR frequencies $f_{\text{FMR}} = (\gamma/2\pi) \sqrt{(B_{\text{e,x}}+B_a)(B_{\text{e,x}}+B_a+B_d)}$ (black curve) closely match for fields applied along the easy axis ($\vu{x}$ direction). The relative error is below 0.003\% for all points.

Figure~\ref{fmrG}(b) presents simulation results for fields applied along the intermediate ($\vu{y}$) axis.  
For $B_{\text{e,y}} < B_a$, the theoretical FMR frequency is given by $f_{\text{FMR}} = (\gamma/2\pi) \sqrt{(B_{\text{e,y}} - B_a)(B_{\text{e,y}} + B_d)}$, and for $B_{\text{e,y}} > B_a$, it is equal to $f_{\text{FMR}} = (\gamma/2\pi) \sqrt{(B_a^2 - B_{\text{e,y}}^2)(B_a+B_d)/B_a}$.  
The circuit model accurately reproduces the mode transition at $B_{\text{e,y}} = B_a = 0.2$~T, with relative errors below 0.02\% for all data points.

Similar results were also obtained for an applied field along the hard-axis ($\vu{z}$), with a relative error below 0.0025\% for all points. 

The agreement between simulated FMR frequencies and theoretical values confirms the accuracy of the equivalent circuit.
The extremely low error values demonstrate the validity of the circuit model for external magnetic fields oriented along the $\vu{x}$, $\vu{y}$, and $\vu{z}$ axes, including cases involving anisotropy and mode transitions (e.g., when $B_{e,y}=B_a$).
These results indicate that the equivalent circuit accurately represents both the conservative (precessional) and damping components of the LLGS equation.

\subsection{Spin-Torque Nano-Oscillator}

STNOs are nanoscale devices that use spin-transfer torque to generate self-sustained oscillations in the free layer magnetization, which produce a time-varying resistance in the MTJ {[}Slavin 2009{]}. 
Here, we demonstrate that the equivalent circuit model accurately reproduces large-amplitude oscillations in a current-driven MTJ.

\begin{figure}[t]
\centerline{\includegraphics{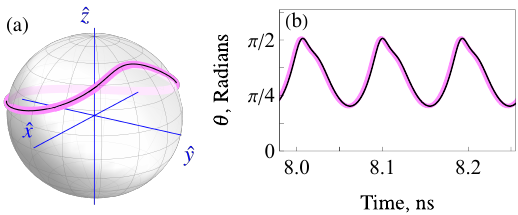}}
\caption{Comparison of STNO dynamics obtained from LTspice and direct simulation of the LLGS equation.  
(a) Simulated trajectory of the free layer magnetization under a 0.33~mA applied current.  
(b) Time evolution of the polar angle $\theta$.  
Black curves correspond to direct numerical simulation of the LLGS equation, while pink curves represent LTspice simulations.
}
\label{stno}
\end{figure}

Figure~\ref{stno}(a) shows the trajectory of the free layer magnetization, $\mathbf{m}$, simulated in LTspice with $V_\mathrm{mtj} = 0.2475$ V, $R_\mathrm{P} = 500~\Omega$, $R_\mathrm{AP} = 1500~\Omega$, spin polarization vector oriented at $\theta_p = 85^\circ$ and $\phi_p = 190^\circ$, Gilbert damping constant $\alpha_G = 0.01$, anisotropy field of $B_a = 0.2$~T, and external magnetic field components $B_{e,x} = -0.1$~T, $B_{e,y} = -0.15$~T, and $B_{e,z} = 0.8$~T.
The free layer magnetization undergoes sustained precessional motion at approximately 14~GHz. 
The trajectory has non-uniform shape due to the combined effects of the effective field orientation and STT orientation and $\theta_p = 85^\circ$.

Figures~\ref{stno}(b) presents the time evolution of the polar angle, $\theta$ and demonstrates that LTspice model produces results identical to LLGS equation. 
This close agreement demonstrates that the equivalent circuit accurately models all three components of the LLGS equation: the conservative precession term, the damping term, and the spin-transfer anti-damping torque.
The simulations specifically employed unconventional external magnetic fields and polarization angles designed to produce non-uniform precession; despite this complexity, the equivalent circuit performed well.
These results may find application in integrated STNO-based systems {[}Litvinenko 2022{]}, where accurate circuit-level simulation of complex magnetization dynamics would be a valuable tool for designing complete circuit implementations.

\subsection{Field-Induced Switching}

In Fig.~\ref{switchin}(a) we compare LTspice and LLGS simulations for field-induced switching of an MTJ.
In this simulation, the system was subjected to a time-varying external magnetic field applied for 0.058~ns at intervals of 1.0~ns, with the field turned off otherwise. 
When active, the field had an amplitude of 0.4~T and was oriented in the $xy$ plane at an angle $15^\circ$ (green shading in Fig.~\ref{switchin}(a)) or $195^\circ$ (blue shading) to MTJ easy axis $\vu{x}$. Other simulation parameters are $R_\mathrm{P} = 500~\Omega$, $R_\mathrm{AP} = 1500~\Omega$, $B_a = 0.1$~T, and $\alpha_G = 0.05$.

The strong agreement between LTspice and LLGS results, including accurate depiction of post-switching ``ringing'', confirms the accuracy of the equivalent circuit model for large-angle ($180^\circ$) magnetization rotation.

The previous two subsections demonstrated that the equivalent circuit functions accurately in stationary regimes, where magnetic fields and currents remain static while the output evolves dynamically.
This section has shown that the circuit also performs reliably under time-varying magnetic fields.
The example of field-induced switching was selected for its familiarity to the MRAM community, but additional tests (not shown) included various time-dependent fields, such as sinusoidal and nonlinear impulses.
In every case, the circuit produced results that closely matched macrospin simulations.
Moreover, these simulations were straightforward to configure and fast to execute using the equivalent circuit in LTspice.

\subsection{Current-Induced Switching in a 1T-1MTJ Cell}

\begin{figure}[t]
\centerline{\includegraphics{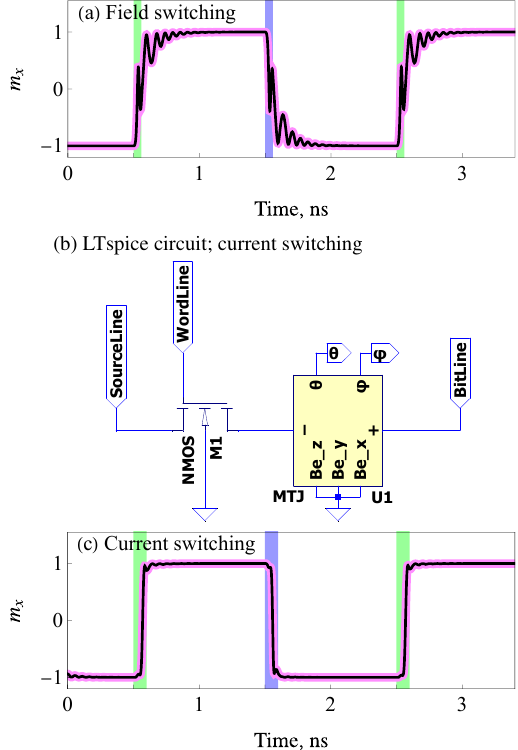}}
\caption{(a) Field-induced switching dynamics. 
(b) Schematic of a 1T-1MTJ cell in LTspice, with the MTJ component highlighted by a yellow box. 
(c) Current-induced switching dynamics in a 1T-1MTJ cell. 
In (a) and (c), pink curves show LTspice simulations while black lines are results of direct numerical simulation of the LLGS equation.
}
\label{switchin}
\end{figure}

To illustrate the practical application of the equivalent circuit model, the MTJ was incorporated into a 1T-1MTJ memory cell and simulated using LTspice. 
The schematic of this memory cell is shown in Fig.~\ref{switchin}(b), and is as previously described in Fig.~\ref{set}(d). 
A 1T-1MTJ cell comprises a single transistor in series with an MTJ, forming a compact and scalable memory element suitable for MRAM architectures {[}Guo 2015{]}. 
In this design, the wordline enables writing, while the bitline and source line supply the required voltages for read and write operations.
Current-induced switching occurs when an electric current reverses the magnetization of the free layer, toggling the MTJ between low- and high-resistance states.

To evaluate the circuit model accuracy, current-induced switching was simulated in a 1T-1MTJ cell and compared with direct LLGS simulation (see Fig.~\ref{switchin}(c)). 
The close agreement confirms the model faithfully reproduces current-induced switching dynamics.

In these simulations, a 5~V square wave with a period of 2 ns was applied alternately to the bitline and source line; the sourceline was at 5~V from 0 to 1~ns, and the bitline was at 5~V from 1 to 2~ns. 
Additionally, a 5~V square impulse lasting 0.1~ns was applied to the wordline at times 0.5 ns, 1.5 ns, and 2.5 ns. 
These applied voltage waveforms resulted in MTJ currents of approximately 9 mA at 0.5 ns and 2.5 ns (green shading) in Fig.~\ref{switchin}(c)) and about –7 mA at 1.5 ns (blue shading).
The spin polarization vector $\mathbf{p}$ was oriented in the $xy$ plane at $15^\circ$ relative to the $\vu{x}$ axis. 
Other simulation parameters were the same as in the field switching example.

This section demonstrates the potential of the equivalent circuit model for simulating hybrid circuits with MTJs and conventional electronic components. 
Although a relatively simple 1T-1MTJ cell was chosen for the example, the equivalent circuit is applicable to circuits of any complexity.
The equivalent circuit has been integrated into a single LTspice part, and readers are invited to download and incorporate this part into their own simulations {[}MTJ Spice Models{]}.

\section{Conclusion}

In this work, we presented a physics-based equivalent circuit model for MTJs that accurately describes both magnetization dynamics and electrical characteristics. 
The proposed circuit model was validated against direct numerical solutions of the LLGS equation across multiple operating regimes, including FMR, STNO, field- and current-induced switching. 
LTspice simulations demonstrated agreement with numerical results, confirming the robustness and fidelity of the model.

All components of the equivalent circuit have a clear physical origin, and, as a result, the proposed model can be easily modified to describe more complicated spintronic devices, such as, e.g., MTJs with multiple free or fixed layers.

This work provides a validated physics-based equivalent circuit model that accurately captures macrospin magnetization dynamics, offering an efficient framework for circuit-level analysis of MTJ devices.
The results presented here support the development of next-generation non-volatile memory and advanced computing technologies based on spintronic elements.

\end{document}